\documentclass[aps,prd,nopacs,twocolumn]{revtex4-1}

\usepackage{amsmath,amssymb,graphicx}
\usepackage[T1]{fontenc}
\usepackage[utf8]{inputenc}
\usepackage{graphicx}
\usepackage{color}
\definecolor{dgreen}{rgb}{0,0.7,0}
\usepackage[unicode=true,pdfusetitle,
 bookmarks=true,bookmarksnumbered=false,bookmarksopen=false,
 breaklinks=false,pdfborder={0 0 0},backref=false,colorlinks=true]
 {hyperref}
\hypersetup{
 citecolor=dgreen,filecolor=dgreen,linkcolor=dgreen,urlcolor=dgreen}

\usepackage{placeins}

\newcommand{\be}{\begin{equation}}
\newcommand{\ee}{\end{equation}}
\newcommand{\bea}{\begin{eqnarray}}
\newcommand{\eea}{\end{eqnarray}}
\newcommand{\bean}{\begin{eqnarray*}}
\newcommand{\eean}{\end{eqnarray*}}

\newcommand{\HH}{{\cal H}}

\newcommand{\NN}{{\cal N}}

\newcommand{\de}{\delta}
\newcommand{\De}{\Delta}

\newcommand{\ga}{\gamma}

\newcommand{\La}{\Lambda}

\newcommand{\Om}{\Omega}
\newcommand{\om}{\omega}
\newcommand{\si}{\sigma}

\newcommand{\lsim}{\stackrel{<}{\sim}}

\def\id{{\rm 1\kern -2.5pt I}}

\begin{document}

\title{Detecting the cosmological neutrino background in the CMB}
\author{Elena Sellentin}
\affiliation{Institut für Theoretische Physik, Ruprecht-Karls-Universität Heidelberg, Philosophenweg 16, 69120 Heidelberg, Germany}
\author{Ruth Durrer}
\affiliation{D\'epartement de Physique Th\'eorique \& Center for Astroparticle Physics,
Universit\'e de Gen\`eve, Quai E.\ Ansermet 24, CH-1211 Gen\`eve 4, Switzerland}

\date{\today}

\begin{abstract}
Three relativistic particles in addition to the photon are detected in the cosmic microwave background (CMB). In the standard model of cosmology, these are interpreted as the three neutrino species. However, at the time of CMB-decoupling, neutrinos are not only relativistic  but they are also free streaming. Here, we investigate, whether the CMB is sensitive to this defining feature of neutrinos, or whether the CMB data allow to replace neutrinos with a relativistic fluid.
We show that free streaming particles are preferred over a relativistic perfect fluid with $\De\chi^2\simeq 21$. We then study the possibility to replace the neutrinos by a viscous fluid and find that also a relativistic viscous fluid with either the standard values $c_{\rm eff}^2=c_{\rm vis}^2=1/3$ or best fit values for  $c_{\rm eff}^2$ and $c_{\rm vis}^2$ has $\De\chi^2\simeq 20$ and thus cannot provide a good fit to present CMB data either. 
\end{abstract}


\maketitle

\section{Introduction}
The cosmic microwave background (CMB) is the most precious cosmological dataset which we analyse to determine the content of our Universe. Alone and in combination with other data it has been used to infer that our Universe is presently dominated by dark energy which may be in the form of a cosmological constant $\La$ contributing a density parameter of $\Om_\La \simeq 0.7$, and pressure less matter which is dominated by cold dark matter with $\om_m=\Om_mh^2=\Om_{\rm cdm}h^2+\Om_bh^2=\om_{\rm cdm}+\om_b\simeq 0.14$ where the contribution from baryons is $\om_b=\Om_bh^2 \simeq 0.022$, see \cite{2015arXiv150201589P} for the latest values. Here $h =H_0/(100$km/s/Mpc) and $H_0$ is the present Hubble parameter. Allowing for generous uncertainties which include the Planck value~\cite{2015arXiv150201589P} as well as local measurements~\cite{Riess:2011} one may allow the range $h\simeq (0.7\pm 0.05)$ for the factor $h$ in which the uncertainty of the value of the present Hubble parameter is  absorbed. (We shall, however not assume any prior for $h$ in our MCMC analysis.)

Furthermore,  there are the photons which make up the CMB
and which contribute $\Om_\ga h^2 =2.48\times 10^{-5}$ and there are cosmic neutrinos. In the standard model of 3 massless neutrino species, they contribute a density parameter of $\Om_\nu h^2 =1.69\times 10^{-5}$. Taking into account neutrino masses, in the minimal model with normal hierarchy and with a maximal neutrino mass of 0.056eV~\cite{Forero:2014bxa},  one obtains $\Om_\nu h^2 \simeq 0.5\times 10^{-3}$.

These are very small numbers. Nevertheless, during the radiation dominated epoch at temperatures above about $1$eV, neutrinos and photons are the dominant constituents of the Universe, and the neutrinos contribute a fraction $f_{\rm rad}=\Om_\nu/(\Om_\ga +\Om_\nu)\simeq 0.4$ to the total energy density of the Universe. At recombination, $z_{\rm dec}\simeq 1100$ they still contribute 
$$f_{\rm dec}\simeq \Om_\nu/\left(\Om_\ga +\Om_\nu +\Om_m/(1+z_{\rm dec})\right) \simeq 0.1 \,, $$ 
i.e.,  10\% to the total energy density of the Universe.

The first indication that  cosmic neutrinos are really present in the Universe in thermal abundance came from nucleosynthesis calculations. The abundance of primordial helium-4 is very sensitive to the expansion rate at temperature $T_{\rm nuc}\simeq 0.08$MeV, which is determined via the Friedman equation by the energy density of the Universe. At this temperature the energy density is dominated by photons and neutrinos. The observed helium-4 abundance requires $N_{\rm eff}\simeq 3\pm 1$ species of neutrinos~\cite{Steigman:2012ve}. Somewhat more stringent results have been obtained  from the recent Planck data, $N_{\rm eff}=2.99\pm 0.4$, see~\cite{2015arXiv150201589P}. 

However, the nucleosynthesis results only require a relativistic component with the given energy density 
in order to provide the correct background expansion. But  neutrinos are not only relativistic in the early universe, they possess additional particle properties to which the background expansion alone is insensitive. Especially, neutrinos are collisionless below $T_\nu \simeq 1.4$MeV corresponding to the  redshift $z_\nu \simeq 10^{10}$ where they decouple from the cosmic fluid. The CMB data is not only sensitive to  the presence of relativistic components in addition to photons via their contribution to the background, but it also allows us to  study their perturbations which are sensitive to additional particle properties of this relativistic species which we can then compare with those expected from true  neutrinos.

In the CMB, neutrinos are usually modeled as collisionless particles in order to mimic  the neutrino free streaming -- but is CMB data really sensitive to this property or could we also fit it with a relativistic fluid?

This is the question we  address in this work. The CMB detects  three relativistic species apart from the photons. We will refer to these as 'neutrinos' for brevity, although at first, it is not clear that they are the standard model neutrinos. We shall assume an agnostic point of view and not choose any model of neutrino interaction. We just study whether a perfect fluid or a viscous fluid of relativistic particles can fit the present CMB data. More detailed studies of constraints of neutrino properties with cosmological data, where the latter are given by concrete physical models and where not only CMB but also large scale structure data is considered can be found in Refs.~\cite{Basboll:2008fx,Archidiacono:2013dua}.
Another model for cosmic neutrinos which we shall call the 'viscous free streaming model'
has  been studied before~\cite{Hu:1998kj,Hu:1998tk,Trotta:2004ty,Archidiacono:2011gq,Gerbino:2013ova,2014arXiv1412.5948A}, however, we shall argue below that this model is not a perfect nor a viscous fluid. The model adopted in these works seems to us less natural but it is certainly complementary to the present analysis.

We first compare the standard CMB-anisotropy calculation with a computation where the relativistic particles are modeled as a perfect fluid. Awaiting the upcoming Planck-data release, we compare our models to the Planck-2013 data with WMAP-polarization.
We use the package Monte Python~\cite{2013JCAP...02..001A}. We use standard cosmological parameters $\omega_b, \omega_{\rm cdm}, h, n_s,A_s, z_{\rm reio}$ along with the Planck nuisance parameters, see~\cite{Ade:2013zuv}.

We fix the background density of the  neutrinos to the Planck-2013 bestfit since the presence of relativistic particles has been accurately measured and we just want to investigate how precisely the particle properties of the detected relativistic particles can be measured. This  means we keep $N_{\rm eff}$ fixed. The primordial helium fraction is then determined by the value of $\omega_b$.  We have checked that allowing the mass of the  neutrinos to vary, does not make any difference to our results. Therefore, for the sake of simplicity, we show the results for one massive eigenstate with the close to minimal mass of $0.06$ eV, the other two eigenstates are treated as massless. We do not adopt any additional priors. Allowing also $N_{\rm eff}$ and with it the primordial helium fraction to vary as a function of $N_{\rm eff}$, does not alter any of the following results. It leads of course to larger error bars on the other parameters. We shall briefly comment on it in the Appendix.

We find that treating neutrinos as collisionless particles fits the data significantly better than a simple relativistic perfect fluid. 

Next we show that neutrinos can also not be modeled as a viscous fluid. 
We also compare our results with the approach which is found in previous literature~\cite{Trotta:2004ty,Archidiacono:2011gq,Gerbino:2013ova} which we explain below.

In the next section we describe our calculations and show the result. In Section~\ref{s:con} we discuss our findings and conclude.

\section{Neutrinos in the CMB}
In standard CMB computations one assumes that neutrinos are massless, free streaming particles and one solves the Liouville equation for them, see, e.g.,~\cite{mybook}. 
\bea
\dot \NN_0 +k\NN_1&=& 0 \,,\label{e:n0}\\
\dot \NN_1 +\frac{k}{3}\left[2\NN_2-\NN_0\right] &=&\frac{k}{3}(\Phi+\Psi_) \,,\label{e:n1}\\
\dot \NN_\ell +\frac{k}{2\ell+1}\left[(\ell+1)\NN_{\ell+1}-\ell \NN_{\ell-1}\right] &=&
0 \,,\quad \ell>1\,.\label{e:nl}\eea
Here $\NN_\ell$ is the $\ell$th multipole moment of the energy integrated neutrino distribution function in Fourier space and $k$ is the wave number, while $\Phi$ and $\Psi$ are the Bardeen potentials.
The moments 0 to 2 are related to the neutrino density perturbation, $\de_\nu$, the potential of the velocity perturbation, $V_\nu$,and the anisotropic stress, $\Pi_\nu$, in longitudinal gauge by
\bea
\de_\nu &=& 4(\NN_0 +\Phi) \,,\label{e:nf0}\\
V_\nu &=& 3\NN_1 \,,\label{e:nf1}\\
\Pi_\nu &=& 12\NN_2 \label{e:nf2} \,.
\eea 
One truly only needs these first three moments of the distribution function since only they enter the energy momentum tensor which  couples to the gravitational field and affects the evolution of the CMB photons. Nevertheless, in the Liouville equation each mode $\NN_\ell$ is coupled by free streaming to $\NN_{\ell+1}$ and $\NN_{\ell-1}$ and therefore to obtain $\NN_0,~\NN_1$ and $\NN_2$ with sufficient precision one usually solves the neutrino hierarchy up to $\ell_{\max}^\nu \sim 10$ -- 20 in order to minimise problems from so called numerical 'reflections'. 

Below we shall consider  treating neutrinos as a relativistic perfect fluid, which corresponds to cutting the hierarchy at $\ell_{\max}^\nu=1$ or as a relativistic viscous fluid, which corresponds to cutting the hierarchy at $\ell_{\max}^\nu=2$. We shall also consider a viscous fluid with arbitrary sound speed  $c_{\rm eff}^2$ and viscosity $c_{\rm vis}^2$. This corresponds to  cutting the hierarchy at $\ell_{\max}^\nu=2$ and replacing  (\ref{e:n0}) to (\ref{e:nl}) by the following system of equations, see~\cite{Hu:1998kj,Hu:1998tk,Trotta:2004ty}:
\bea
\dot \NN_0 +k\NN_1&=&\HH(1-3c_{\rm eff}^2)\NN_0 \label{e:nc0}\\
\dot \NN_1 +\frac{k}{3}\left[2\NN_2-3c_{\rm eff}^2\NN_0\right] &=& \nonumber\\
&& \hspace*{-3.4cm}-\HH(1-3c_{\rm eff}^2)\NN_1 +\frac{k}{3}(3c_{\rm eff}^2\Phi+\Psi ) \label{e:nc1}\\
\dot \NN_2 +k\left[\frac{3}{5}\NN_3 -3c_{\rm vis}^2\frac{2}{5}\NN_1\right] &=&0 \label{e:nc2}\eea
\be
\dot \NN_\ell +\frac{k}{2\ell+1}\left[(\ell+1)\NN_{\ell+1}-\ell \NN_{\ell-1}\right] =
0 \,,\quad \ell>2\,.  \label{e:ncl}
\ee
In addition, the fact that the perturbations have to be evaluated in the rest frame of the fluid, leads to subtle changes of Eqs.~(\ref{e:nc0}-\ref{e:nc2})  as described in \citep{2014arXiv1412.5948A}. We include these for our modeling of the massive and massless neutrinos. In our viscous fluid model 
we differ from the treatment in the above mentioned papers by cutting the hierarchy at $\ell_{\max}^\nu=2$ in the and we consistently also set $\NN_3\equiv 0$.
 
\begin{figure}[ht!]
 \centering
   {\includegraphics[width=9cm]{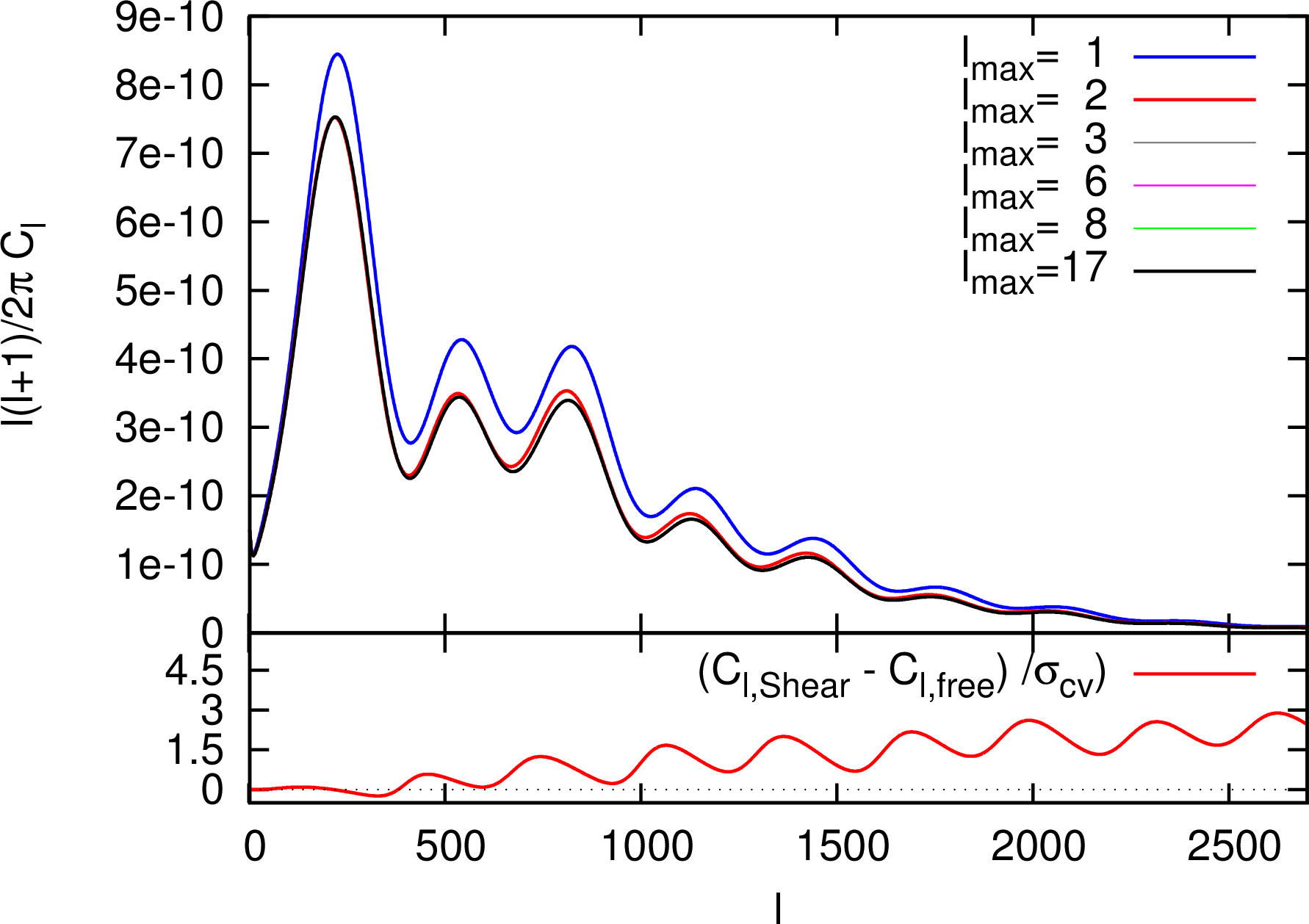}}
   \caption{\label{f:comp}The temperature anisotropy spectra at fixed cosmological parameters for different values of the cutoff in the neutrino hierarchy, $\ell_{\max}^\nu=1,2,3,6,8,17$. For $\ell_{\max}^\nu>2$ the result changes very little. In the bottom panel, the difference between $C_\ell$ for the viscous neutrino fluid ($\ell_{\max}^\nu=2$) and free streaming neutrinos, is compared to  cosmic variance, which roughly corresponds to the Planck error out to $\ell\simeq 2000$. For low $\ell$ cosmic variance does not allow to discriminate between the viscous fluid model and free streaming neutrinos, but for high $\ell$ the difference between these two models is up to three times larger than  cosmic variance $\sigma_{cv} = C^{\rm free}_\ell \cdot \sqrt{2/(2\ell +1)f_{sky}}$.  (Note that here $\ell_{\max}^\nu$ denotes the maximal neutrino multipole while $\ell$ refers to the CMB multipoles.)}
\end{figure}

In Fig.~\ref{f:comp}, we show the CMB anisotropy power spectrum for fixed cosmological parameters by modeling the neutrino hierarchy up to $\ell_{\max}^\nu$. 
Already for $\ell_{\max}^\nu=2$
the difference between the standard calculation setting $\ell_{\max}^\nu=17$
becomes very small. Nevertheless, as is visible from the lower panel, for a cosmic variance limited experiment, like Planck for $\ell\lsim 2000$, this difference for $\ell_{\max}^\nu=2$ is highly significant.

\begin{figure}[ht!]
 \centering
   {\includegraphics[width=8cm]{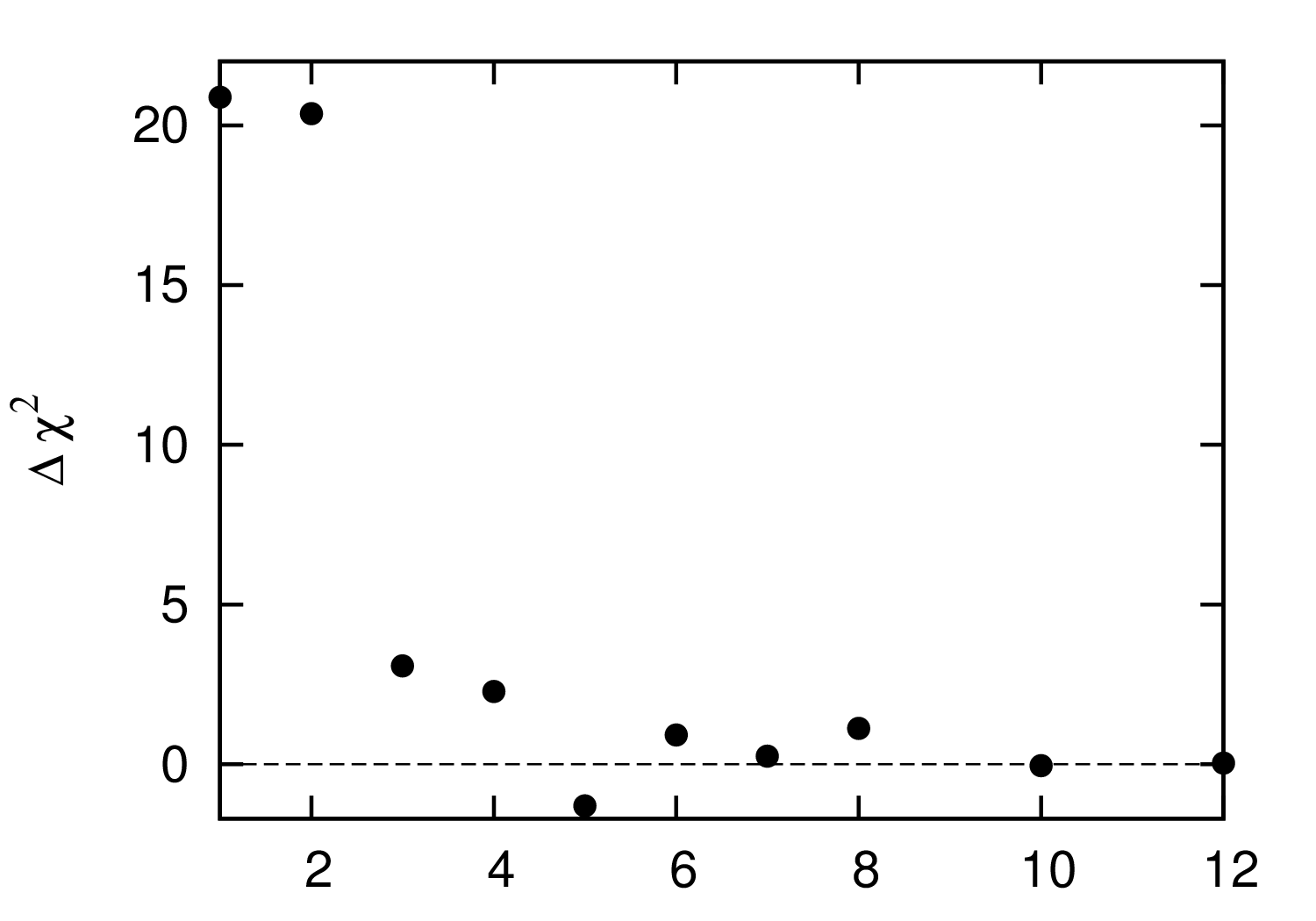}}
   \caption{\label{f:Lmax}The improvement of the fit, $\Delta \chi^2 = \chi^2(\ell_{\max}^\nu) - \chi^2 (\ell_{\max}^\nu = 17)$, as a function of the maximal considered neutrino  multipole $\ell_{\max}^\nu$ in the Liouville equations for the neutrinos. free streaming neutrinos correspond to $\Delta \chi^2 = 0$.}
\end{figure}

In Fig.~\ref{f:Lmax}, we show how the fit to the data improves as a function of the maximally allowed neutrino multipole: truncating at $\ell_{\max}^\nu = 1,2$ leads to the deteriorated fits of the ideal and relativistic viscous fluid. Truncating at $\ell_{\max}^\nu = 5$ leads actually to a slightly better fit than solving the Boltzmann hierarchy up to $\ell_{\max}^\nu = 17$. At the same time, it is evident from Fig.~\ref{f:triangle} below that the best fit values and the $1\sigma$ contours of the cosmic parameters do not change when cutting the hierarchy anywhere between $\ell_{\max}^\nu = 4$ and $\ell_{\max}^\nu = 17$. The negative $\Delta \chi^2 = -1.29$ for $\ell_{\max}^\nu=5$ therefore does not seem to stem from physics but might be due to numerical inaccuracies and to the modeling of the experimental uncertainties (see also the discussion about model 'evidence' below). But even if truncating at different $\ell_{\max}^\nu$ may lead to typical changes in $\Delta \chi^2$ on the order of unity, the conclusion that the ideal and the standard viscous fluid are worse fits than free streaming neutrinos which we shall draw below, remains valid since their $\Delta \chi^2$ is much higher.

\begin{figure}[ht!]
 \centering
   {\includegraphics[width=8.5cm]{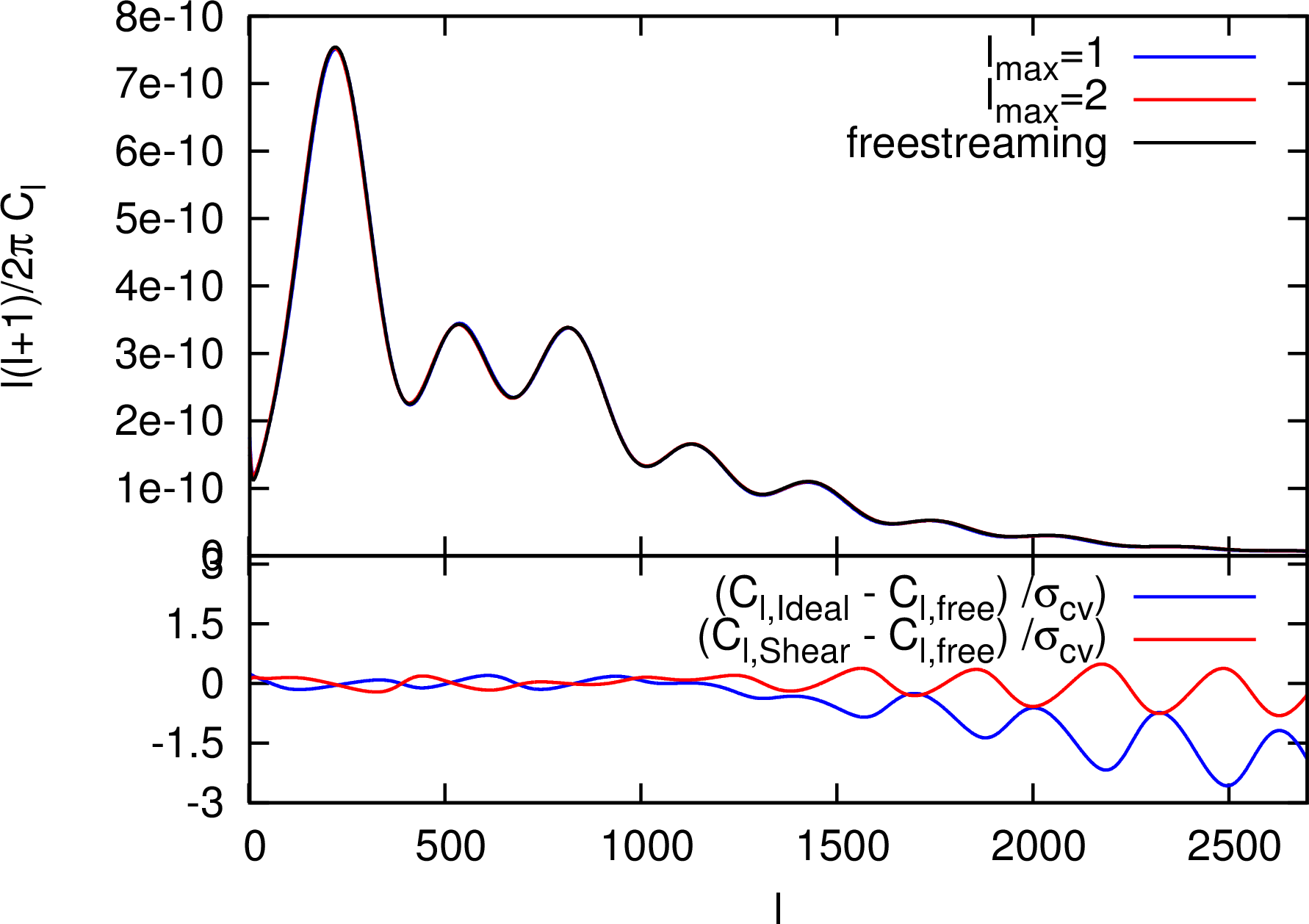}}
   \caption{\label{f:spec}The temperature anisotropy spectra for best fit parameters modeling neutrinos as a perfect fluid (blue), a relativistic viscous fluid (red) and standard free streaming neutrinos (black) are shown. The difference of the best fit spectra is not visible by eye. However, from the plot at the bottom which shows the difference in units of the cosmic variance error it is clear, that the Planck experiment can distinguish the spectra.}
\end{figure}

\begin{figure*}[ht!]
 \centering
   {\includegraphics[width=17cm]{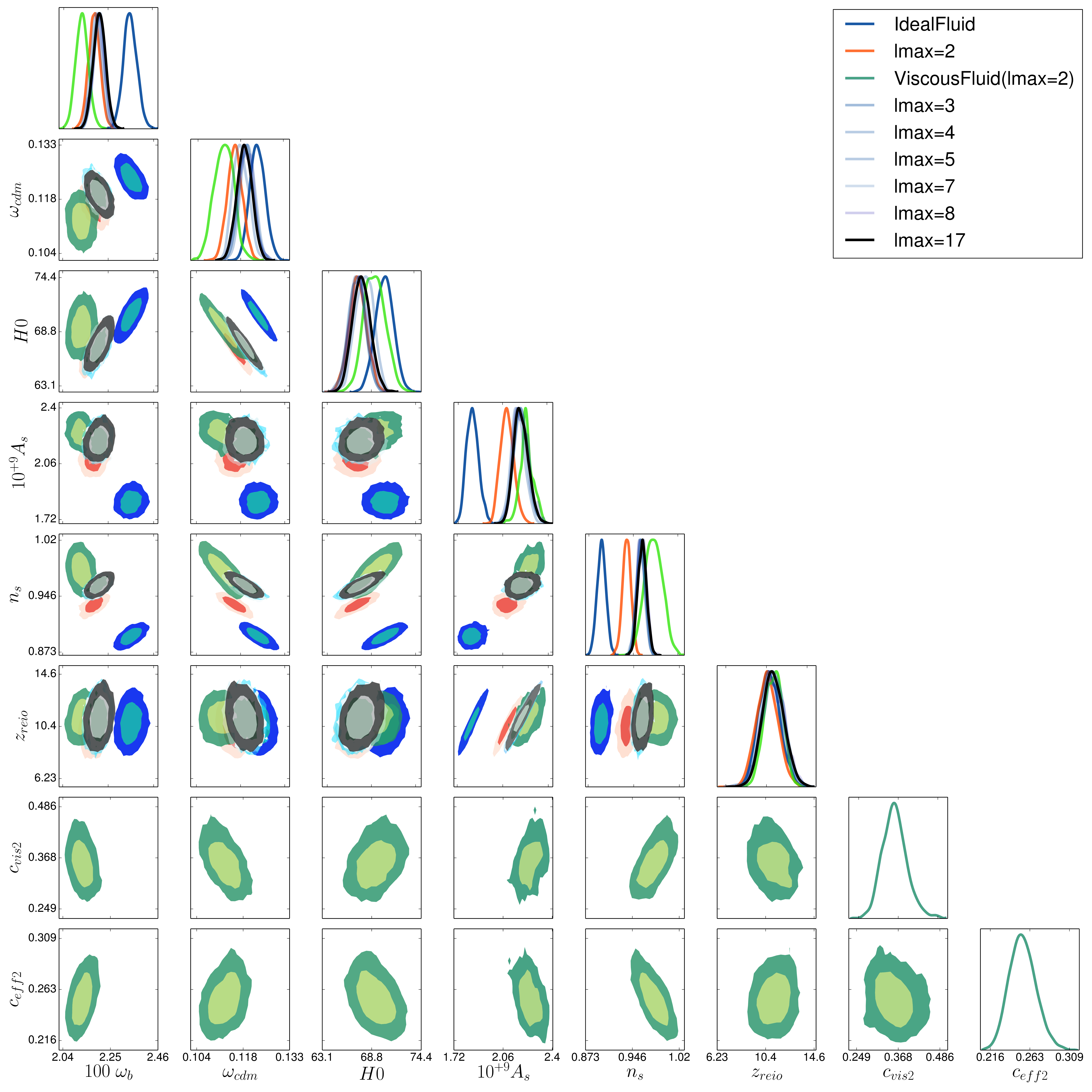}}
   \caption{\label{f:triangle} The best fit parameters for neutrinos modeled as a perfect fluid (blue),  as a relativistic viscous fluid  (orange),   as a viscous fluid with arbitrary sound speed $c_{\rm eff}^2$ and viscosity $c_{\rm vis}^2$ (green), and as standard free streaming neutrinos (black) are shown.  We also show free streaming neutrinos with different $\ell^\nu_{\max}$ as indicated in the legend  in light blue to grey shades. The best fit values of several parameters for the perfect fluid and the free streaming model  differ significantly.  The best fit values of most parameters for the  viscous fluid and the free streaming model  are similar, they all agree within $1.5\si$ apart from $n_s$ which for the relativistic viscous fluid model differs by more than $2\si$. Truncating the Boltzmann hierarchy for the neutrinos at other maximally allowed $\ell \geq 3$ (indicated in different shades of blue) leads to parameter constraints that are indistinguishable from the standard Planck fit, consistent with Fig.~\ref{f:comp}. Allowing also $N_{\rm eff}$ does not change these results, as can be seen in Fig.~\ref{Neff} in the appendix.}
\end{figure*}

We have investigated whether neutrinos can be modelled by a relativistic perfect or viscous fluid.
For this, we have replaced the massless and massive neutrinos by a (relativistic) perfect fluid or a (relativistic) viscous fluid and run the modified CMB code CLASS~\cite{Blas:2011rf,Audren:2013} in combination with Monte Python~\cite{2013JCAP...02..001A} to find best fit values of the standard cosmological parameters from the Planck data. In Fig~\ref{f:spec} we compare the spectra obtained in this way with the spectrum from free streaming neutrinos and in Fig.~\ref{f:triangle} we show the best fit parameters.  By eye, the curves look identical. But when considering the difference in units of the cosmic variance, it becomes clear that a cosmic variance limited experiment like Planck can measure the small difference.

Not only are several of the cosmological parameters significantly different, see Fig.~\ref{f:triangle}, but the fit is also much worse. The $\De\chi^2$ for both  fluid approximations increases by:
\be\label{e:chi2}
\De\chi^2_{\rm ideal} \approx  21\,, \qquad
\De\chi^2_{\rm visc} \approx  20\,.
\ee

Let us formulate this in terms of the Bayes factor, $K$ (see~\cite{Jef,Kass:1995}) which indicates whether model $M_1$ (for us free streaming neutrinos) is favoured over model $M_2$ (for us either ideal fluid or viscous fluid neutrinos). The Bayes factor is defined as~\cite{Jef}
\be\label{e:Bayes}
K \equiv \frac{P(D|M_1)}{P(D|M_2)} =\frac{P(M_1|D)P(M_2)}{P(M_2|D)P(M_1)}
= \frac{P(M_1|D)}{P(M_2|D)} \,.
\ee
The last equal sign is due to the fact that in our case, both models have the same parameters so the model spaces are identical. The difference is only that in model $1$ the neutrinos are the standard free streaming neutrinos while in model $2$ they are an ideal or a viscous fluid. Since we want to test exactly this hypothesis (or rather we want to see whether the CMB data is sensitive to this hypothesis) we cannot give model $2$ a smaller model probability. We therefore set $P(M_1)=P(M_2)$. Here $P(M|D)$ is the probability of a model given the data while $P(D|M)$ is the probability of the data given the model. The prior, $P(D)$ drops out in the ratio $K$.
 But the $P(M|D)$ are exactly the likelihoods which we determine in our MCMC code, so that 
\be
2\log(K) = \De\chi^2\,.
\ee
According to Ref.~\cite{Kass:1995}, while $-2<2\log(K)<2$ is 'not worth mentioning' a value $2<2\log(K)<6$ can be interpreted as 'positive' but not strong evidence while $10< 2\log(K)$ is 'decisive' evidence (see also Appendix~B in~\cite{Jef}).  First we conclude that all the fluctuations in $\De\chi^2$ for $\ell^\nu_{\max}\ge 4$ are 'not worth mentioning'.
The contrary holds for   $\ell^\nu_{\max}\le 2$, in this case the evidence in favour of the free streaming model is truly 'decisive'. 

This shows that cosmic neutrinos cannot be modelled neither by a relativistic perfect fluid nor by a viscous fluid. Nevertheless, one may be surprised that the ideal fluid model is not much more strongly excluded than the viscous fluid. Also, our value of $\De\chi^2_{\rm ideal}$ is significantly smaller than other values published in the literature, see e.g.~\cite{Archidiacono:2013dua}. One reason for this at first surprising finding is that we include all Planck nuisance parameters in our MCMC analysis and allow for the priors suggested by the Planck collaboration. As an illustration of how this affects the results we show the 1d likelihoods of the nuisance parameters in  the appendix, Fig.~\ref{f:nuisance}. As one can see there, the fluid models prefer different values for some of the nuisance parameters, especially the best fit amplitudes of the inferred kinetic SZ-effect, $A_{ksz}$ and of the cosmic infrared background at 146GHz, $A_{cbi143}$ are very different for the ideal fluid model than for the viscous fluid or free streaming model, but as is evident from Fig.~\ref{f:nuisance}, these parameters are very badly constrained by the data.
The increase in $\De\chi^2_{\rm ideal}$ in Ref.~\cite{Archidiacono:2013dua}, however  is not only due to a less conservative CMB analysis but also to the inclusion of large scale structure data. We avoid the inclusion of large scale structure data here, in order to keep the properties of neutrinos at high energies (i.e. in the early universe) separate from late-time cosmology effects, where properties of neutrinos can also be used to explain dark energy phenomenology~\cite{2013PhRvD..87d3519A}. Furthermore, the topic of this work is whether the free streaming of neutrinos can be detected in the CMB.

To test the importance of the nuisance parameters we have also run a chain where we fixed them to their best fit values in the free streaming model. This, of course, reduces the size of the model space significantly. In this case the increase in $\De\chi^2$ for both, the ideal and the viscous fluid model in comparison to standard LCDM is larger namely
\bea \label{e:chifix}
\De\chi^2_{\rm ideal} \approx  32\,, \qquad
\De\chi^2_{\rm visc} \approx  30 \,,\\
 \mbox{with fixed nuisance parameters.}\nonumber
\eea

However, these increases in $\De\chi^2$ are not trivial to interpret: fixing the nuisance parameters to the best-fit values of the Planck collaboration is a form of including knowledge about which parameter values the current CMB data prefer, when fitted with free streaming
neutrinos. Therefore, the self-consistent values are those given in Eq.~(\ref{e:chi2}). 

Let us also compare this analysis with previous work~\cite{Trotta:2004ty,Archidiacono:2011gq,Gerbino:2013ova} on neutrino clustering properties, where a somewhat different standpoint has been taken. There, eqs. (\ref{e:n0}) to (\ref{e:nl}) are replaced by eqs.~(\ref{e:nc0}) to (\ref{e:ncl}) .
A similar, non-perfect-fluid treatment has already been suggested in Refs.~\cite{Hu:1998kj,Hu:1998tk}. However, eqs.~(\ref{e:nc0}) to (\ref{e:ncl}) describe neither a perfect nor an imperfect fluid since the higher moments, $\ell\ge 3$, are not damped by collisions but evolve like those of free streaming particles. We dubb this mixture model 'viscous free streaming model'.

The advantage of the viscous free streaming model is that it is 'nested' inside the standard model of free streaming neutrinos with two additional parameters which take the values $c_{\rm eff}^2=c_{\rm vis}^2=1/3$ in the standard model and previous work, especially~\cite{Gerbino:2013ova} have found that the preferred values of these parameters are indeed close to the standard relativistic ones.
Nevertheless, the  physical meaning of $c_{\rm eff}$ and $c_{\rm vis}$ remains unclear since only the evolution of the first and second moment but not higher moments are affected by collisions in this model. This seems surprising to us and we are not aware of a physical example which leads to such a behaviour.  Usually, viscosity damps out all higher moments and thereby inhibits free streaming of all higher moments. Nonetheless, it has been found that the viscous free streaming model succeeds in fitting also the latest Planck Data \cite{2015arXiv150201589P}, preferring again the standard values for $c_{\rm eff}^2=c_{\rm vis}^2=1/3$ which represent in fact the only case in which the inconsistency between viscosity and free streaming vanishes. For a more detailed  discussion of how the cutting of $\ell$-modes and the effective fluid parameters $c_{\rm eff}^2$ and $c_{\rm vis}^2$  map to particle properties, see~\citep{2015JCAP...04..016O}.  

Modelling a true viscous fluid however, not only requires the introduction of the new parameters $c_{\rm eff}^2$ and $c_{\rm vis}^2$ but also either cutting  the neutrino hierarchy at $\ell = 2$ or describing the evolution of the higher moments with a collision term, as e.g. in~\cite{Basboll:2008fx}.

In our model of a viscous fluid we set $\NN_\ell\equiv0$ for all $\ell\ge 3$ and fit for $c_{\rm eff}^2$ and $c_{\rm vis}^2$. As we have discussed above, this model with $c_{\rm eff}^2=c_{\rm vis}^2=1/3$, i.e., the relativistic viscous fluid, provides a bad fit to the observed CMB anisotropies. Before concluding that the three relativistic particles in the CMB are indeed free streaming neutrinos, we need, however, to check whether another value of $c_{\rm eff}^2$ and $c_{\rm vis}^2$ might provide a better fit. Introducing two new free parameters, will of course improve the fit, and we find that the difference in $\chi^2$ for the best fit  with respect to the relativistic viscous fluid is: $\De\chi^2 \approx - 0.58$, with the negative sign indicating an improvement of the fit. The standard value of  $c_{\rm eff}^2 = 1/3$  is excluded at $3\sigma$ whereas $c_{\rm vis}^2 = 1/3$ is compatible within one standard deviation. Nevertheless, in this case, the model space of the new model 2 is increased which enhances the Bayes factor in favour of model 1.  A rule of thumb is that each new parameter has to improve $\De\chi^2$ by at least 1 in order to compensate the 'Occam's razor factor' $P(M_2)/P(M_1)$ in Eq.~(\ref{e:Bayes}). Hence the modest improvement of the best fit by $0.58$ after the introduction of two new parameters, leads to the conclusion that these two new parameters are not  justified.

For the viscous free streaming model, we find that the fit improves by $\De\chi^2 \approx - 3.72$ with respect to free streaming neutrinos. This improvement of the fit is somewhat 
stronger than what is usually expected when adding two additional parameters, $\De\chi^2 \approx -2$, but it leads to a $2\log(K)\simeq\De\chi^2 +2 = -0.72$, hence 'not worth mentioning'. 

\begin{figure*}[ht!]
 \centering
   {\includegraphics[width=18cm]{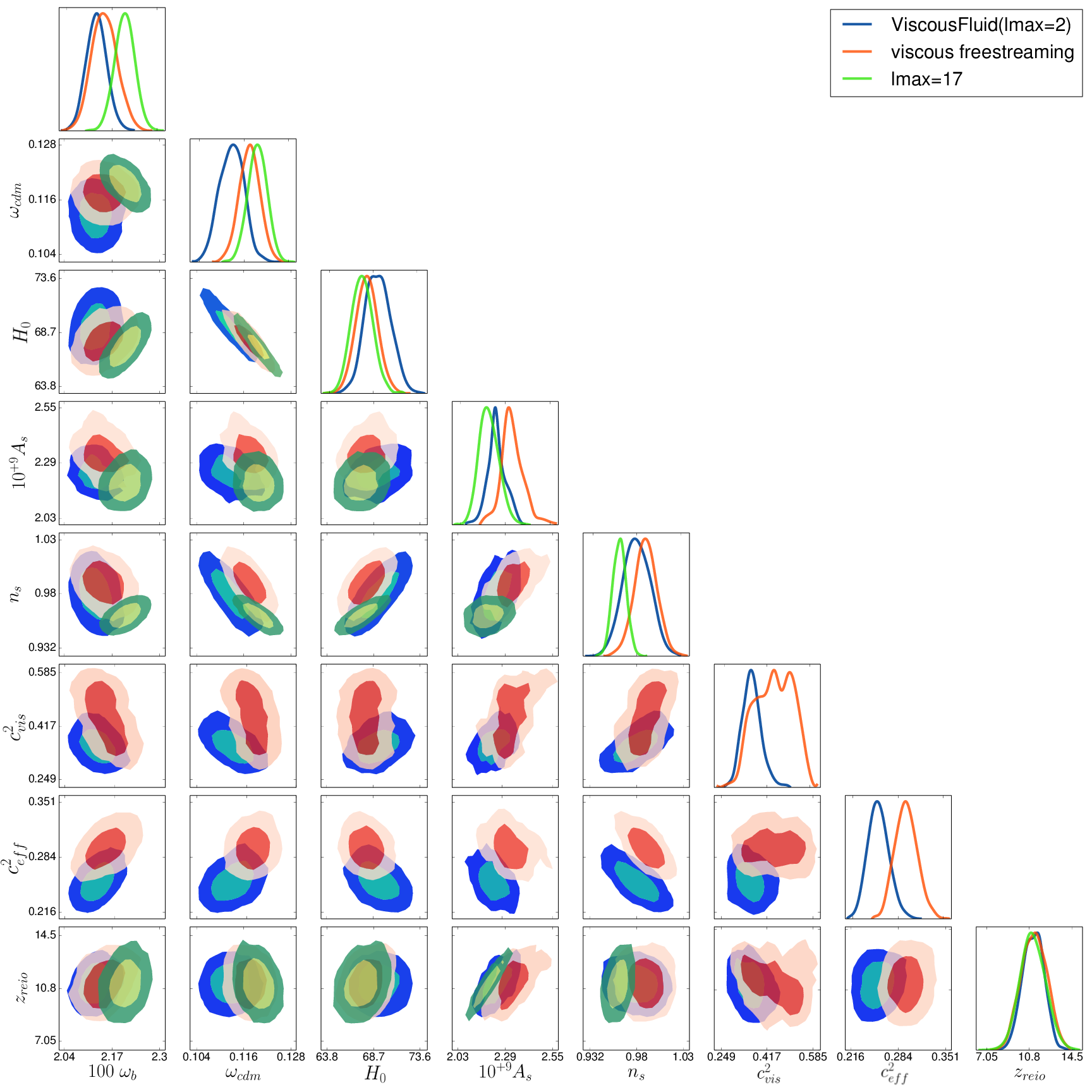}}
   \caption{\label{f:viscousfree streaming}The best fit parameters for neutrinos modeled as a viscous fluid with variable $c_{\rm eff}^2$ and $ c_{\rm vis}^2$  and $\ell^\nu_{max = 2}$ (blue), for neutrinos modeled as a  viscous free streaming fluid with $\ell^\nu_{max} = 17 $(orange),  and for standard free streaming neutrinos  (green) are shown. The best fit values of most parameters for the two different viscosity models  differ by about one standard deviation.}
\end{figure*}

In Fig.~\ref{f:triangle} we have compared the parameter values obtained by replacing neutrinos by a perfect fluid, a relativistic viscous fluid  or by a viscous fluid with arbitrary effective sound speed $c_{\rm eff}^2$ and viscosity, $c_{\rm vis}^2$ with the results for standard neutrinos. 
For completeness, we compare in Fig.~\ref{f:viscousfree streaming}  also the parameter constraints for variable viscosity parameters $c_{\rm eff}^2=c_{\rm vis}^2$ for a viscous free streaming fluid and for a true viscous fluid that cannot build up moments with order higher  than $\ell = 2$.

\section{Conclusions}\label{s:con}
We have studied how neutrinos are detected in the CMB. We have shown that they are not only relevant as additional relativistic degrees of freedom, but  CMB anisotropies and polarisation are also very sensitive to their clustering properties. While the Planck-2013 data is in good agreement with free streaming neutrinos, it cannot be fitted by neutrinos modeled as a relativistic perfect fluid. The best fit model with perfect fluid neutrinos leads to a $\De\chi^2=21$ with respect to the best fit free streaming neutrinos.
Even including anisotropic stress, i.e. allowing for a relativistic viscous
fluid cannot fit the data. The increase in $\chi^2$ with respect to the best fit models with free streaming neutrinos is 
$\De\chi^2 \approx 20 $. Using the evidence scale introduced by Jeffrey~\cite{Jef}, this result can be considered as 'decisive evidence' for neutrino free streaming in the CMB.

The fit can be improved only slightly when allowing arbitrary values for the effective sound speed and the viscosity, $c_{\rm eff}^2$ and $c_{\rm vis}^2$.  But
including these two additional parameters and truncating the neutrino Boltzmann hierarchy at $\ell^\nu_{\max} =2 $ in order to consistently model a viscous fluid, the fit improves by $\De\chi^2=-0.58$ w.r.t. the best relativistic viscous fluid model with fixed values $c_{\rm eff}^2=c_{\rm vis}^2=1/3$.  Therefore, the introduction of these additional parameters is not favoured.

Using observations of the cosmic microwave background anisotropies and polarization we have not only found that there are 3 species of light particles, but we can also infer that these relativistic particles are free streaming. This is a significant additional step towards the detection, albeit indirect, of the cosmological neutrino background. These results are robust under the variation of $N_{\rm eff}$ as an additional parameter and  $N_{\rm eff}$ is found to peak at the standard value of three neutrino species.
\vspace{0.1cm}

\paragraph*{Acknowledgments}
We enjoyed  interesting and helpful discussions with Martin Kunz, Francesco Montanari and Jaspreet Sandhu. We also thank the system administrator Elmar Bittner for maintaining the ITPs computing network.
RD acknowledges financial support from the Swiss NSF. ES acknowledges financial support through the RTG \emph{Particle Physics beyond the Standard Model}, through the DFG fund 1904 and thanks Geneva University for hospitality during part of this work.

\bibliographystyle{apsrev4-1}
\bibliography{lit}

\FloatBarrier
 ~~~

\newpage

\appendix*
\section{Likelihoods of the nuisance parameters, $N_{\rm eff}$}
\begin{figure*}[ht!]
 \centering
   {\includegraphics[width=18cm]{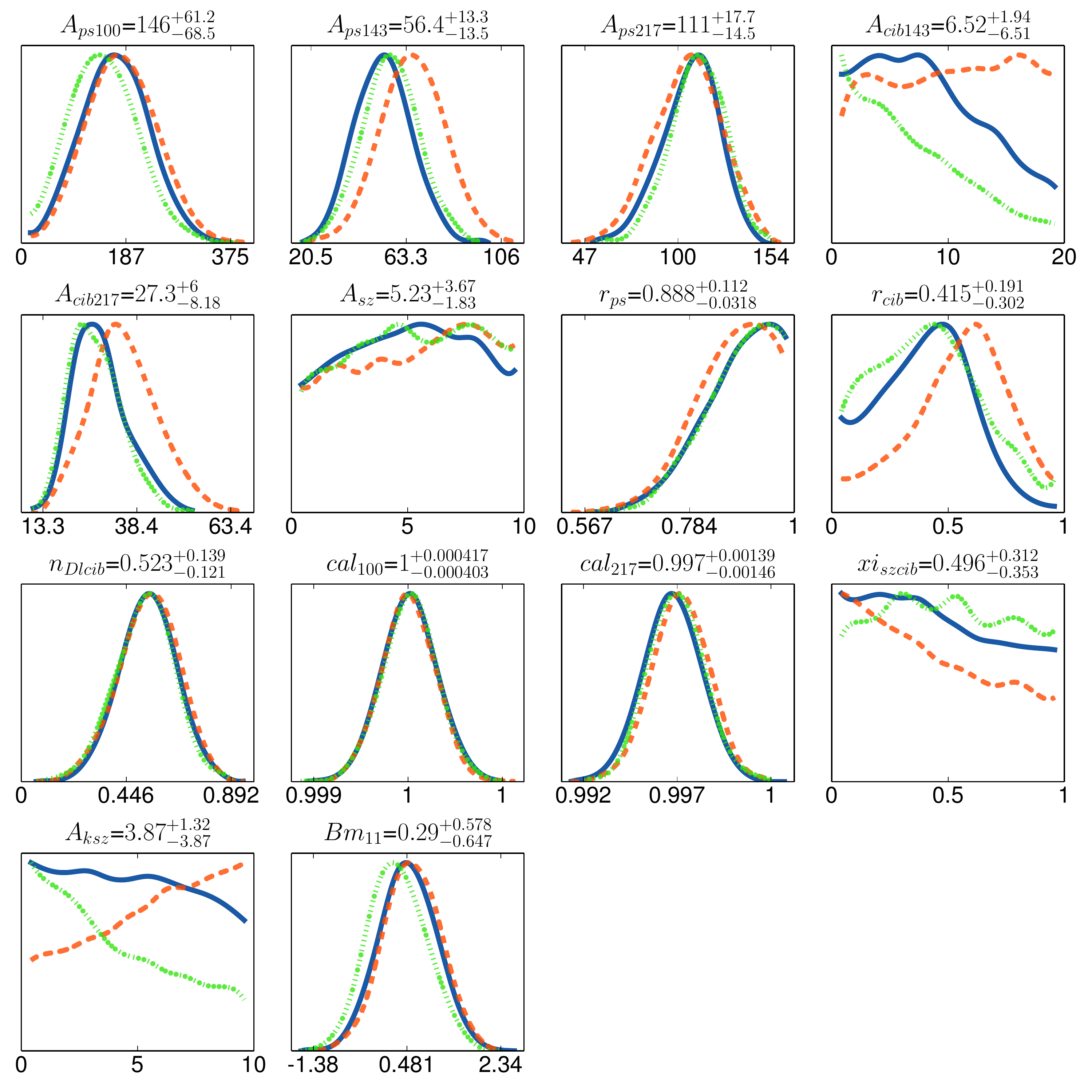}}
   \caption{\label{f:nuisance}The 1d likelihoods of the Planck nuisance parameters  for  free streaming neutrinos (solid, blue), the viscous fluid model (dotted, green) and the ideal fluid model (dashed, orange).  The best fit values indicated on top of each panel are those of the viscous model. Especially the best fit amplitudes of the kinetic SZ-effect, $A_{ksz}$ and of the cosmic infrared background at 146GHz, $A_{cbi146}$ are not well constrained and are very different for the ideal fluid model and for the viscous fluid or free streaming model.}
\end{figure*}

In Fig~\ref{f:nuisance} we show the marginalised 1-parameter distribution of the nuisance parameters used in our analysis of the Planck data. We use the priors as suggested by the Planck Collaboration~\cite{Ade:2013zuv}.  As one sees in the figure, several of these parameters are not well constrained by the data. Also, in the perfect fluid model several nuisance parameters take quite different values than for the viscous fluid or the free streaming model. This allows the perfect fluid model to fit the data not significantly worse than the viscous fluid model. Fixing these parameters  leads to a somewhat larger value of $\De\chi^2_{\rm ideal} -\De\chi^2_{\rm visc} \simeq 2$ (see Eq.~(\ref{e:chifix}).

\begin{figure*}[ht!]
 \centering
   {\includegraphics[width=17cm]{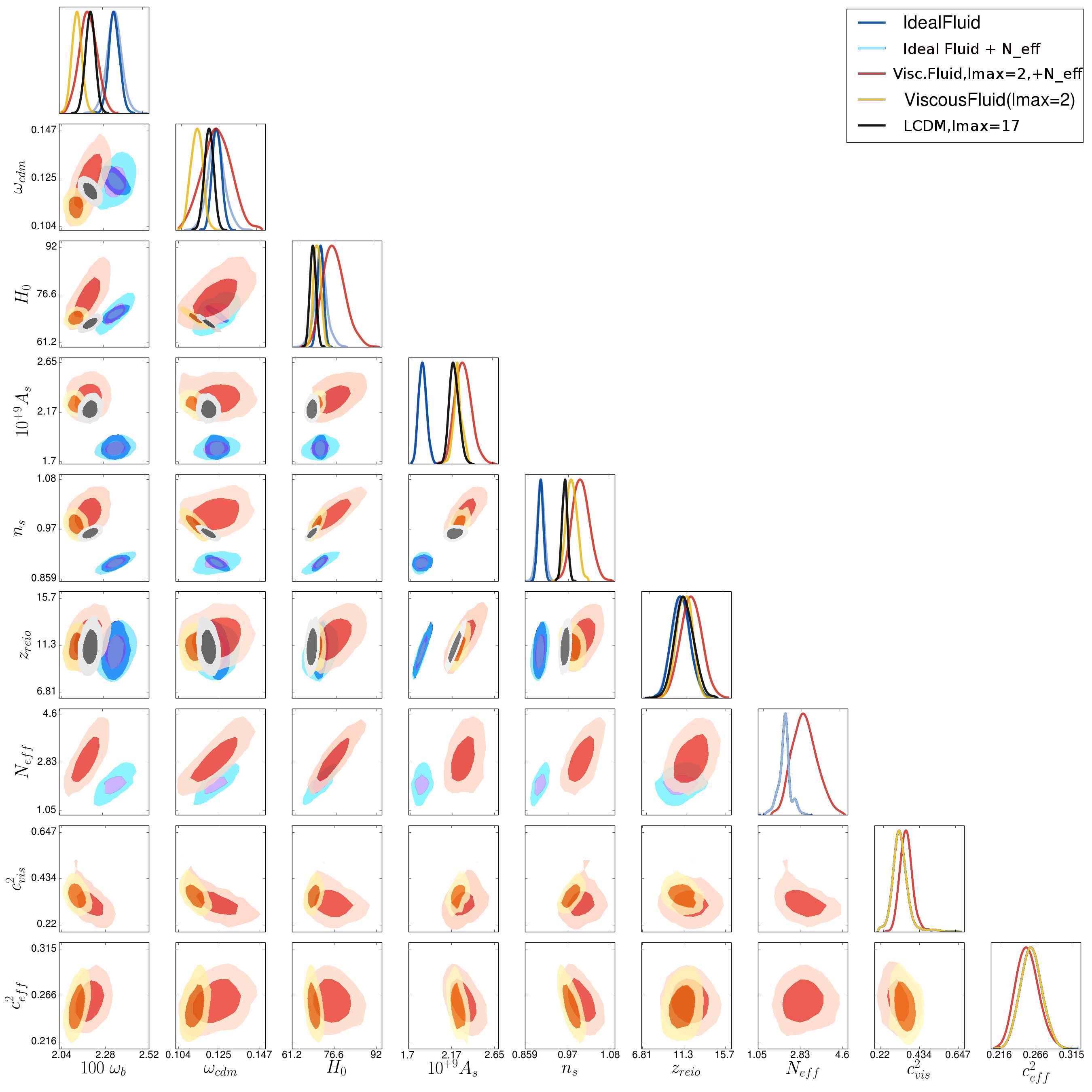}}
   \caption{\label{Neff} Comparing the results of Fig.\ref{f:triangle}, with the results for a varying $N_{\rm eff}$. The inclusion of the parameter $N_{\rm eff}$ does not significantly alter the best fit values of the other cosmological parameters, and $N_{\rm eff}$ itself peaks at around the standard value of three neutrino species. The 1-parameter likelihoods of the ideal fluid model remain virtually unchanged. Those of the viscous fluid model change mainly by widening. The best fit parameters move by less than one standard deviation.}
\end{figure*}

In Fig.~\ref{Neff} we compare the two dimensional likelihoods of both, the ideal fluid model and the viscous fluid model 
 with and without varying $N_{\rm eff}$. Even though the error bars of course increase when including this additional parameter, the main results are unchanged. While the peak value of $N_{\rm eff}$ for the viscous model changes by less than one standard deviation, the ideal fluid would actually prefer a lower $N_{\rm eff}$. Note also that the width of the distribution of $N_{\rm eff}$ in the viscous fluid model is very similar to the standard one, (see~\cite{Ade:2013zuv}, Fig.~21) while for the ideal fluid  model $N_{\rm eff}$ is somewhat more constrained. Nevertheless, the increase in $\De\chi^2$, which is the main point of this study, remains stable.

\end{document}